\documentclass[twoside,centered,14pt,reqno]{article}
\usepackage{lineno,hyperref}
\usepackage{amssymb,latexsym,epsfig,amsmath}
\usepackage{amsthm}
\usepackage{color}
\usepackage{subfig}
\usepackage{rotating}
\usepackage{mathtools}
\usepackage{graphicx}

\usepackage{bm}
\usepackage{multirow}

\usepackage{mathtools}
\usepackage{multirow, longtable}

\newcommand{\be}{\begin{equation}}
\newcommand{\ee}{\end{equation}}

\newcommand{\myfrac}[2]{\displaystyle \frac{#1}{#2}}
\newcommand{\ii}{{\rm i}}

\modulolinenumbers[5]
\begin{document}
%\vskip 0.50truecm
\font\title=cmbx14 scaled\magstep2
\font\bfs=cmbx12 scaled\magstep1
\font\little=cmr10
\begin{center}
{\title {\Large \bf {The Kramers-Kronig relations \\and the analogy between electromagnetic 
\\  and mechanical waves}}}
 \\ [0.25truecm]
{\Large J. Carcione$^{1,2}$,   F.  Mainardi$^{3}$,
 J. Ba$^{1}$, J. Chen$^{1}$}
\\ [0.25truecm]
$\null^1${\little School of Earth Sciences and Engineering, 
Hohai University, Nanjing, China}
\\ [0.25truecm]
$\null^2$ {\little National Institute of Oceanography and Applied Geophysics, Trieste, Italy}
\\ [0.25truecm]
$\null^3$ {\little Department of Physics and Astronomy, University of Bologna, Italy}
%\\  {\little Via Irnerio 46, I-40126 Bologna, Italy}
%\\

\end{center}
%%%%%%%%%%%%5

\begin{abstract}
The important consequence of the Kramers-Kronig  relations (KKrs)
is that dissipative behavior in material media inevitably implies the existence of
dispersion, i.e., a frequency dependence in the constitutive equations. 
Basically, the relations are the frequency-domain expression of causality and 
correspond mathematically to pairs of 
Hilbert transforms.
\\
The relations have many forms and can be obtained with diverse mathematical tools.  Here, two different demonstrations are given in the electromagnetic case, illustrating the eclectic mathematical apparatus available for this purpose. Then, we apply the acoustic (mechanical)-electromagnetic analogy to obtain the elastic versions. 
One major consequence is wave propagation attenuation and pulse spreading, that is, the progressive
widening of a pulse as it propagates through a medium [vacuum seems to be the only ``medium" where this does not
occur (electromagnetic dispersion), while mechanical waves do not propagate]. Therefore, we derive KKrs that relate the wave velocity to the attenuation and quality factors. 
Finally, we discuss the concepts of stability and passivity and provide a novel algorithm to compute the relations numerically by using the fast Fourier transform.

\vspace{0.4cm}

\noindent {\bf Mathematics Subject Classification:} 
%26A33, 33E12, 35C05, 44A10, 60H30										
%26  Real functions: 26A33 Fractional derivatives and integrals
%33 Special functions:33E12 Mittag-Leffler functions and generalizations
%35 Partial differential equations:35C05 Solutions to PDEs in closed form
%44 Integral transforms, operational calculus:44A10 Laplace transform
% 60 Probability theory and stochastic processes:
%60H30 Applications of stochastic analysis (to PDEs, etc.)
%
42A38,74J05,78A40;
%42A38 Fourier and Fourier-Stieltjes transforms and other transforms of Fourier type
%74J05 Linear waves in solid mechanics
%78A40 Waves and radiation in optics and electromagnetic theory
\vspace{0.3cm}

\noindent {\bf Keywords:} 
Kramers-Kronig relations; Fourier transform; Hilbert transform;
Electromagnetic waves; Acoustic waves.

\vspace{0.3truecm} \noindent
{\bf Published in Bulletin of Geophysics and Oceanography, \\ Vol. 63, n. 2, pp. 175-188; 
June 2022, DOI 10.4430/bgo00380}

\end{abstract}

\section{Introduction}

Many scientists contributed to the wave theory, and since the beginnings 
there has been an interplay between the theory of light and mechanical wave propagation. 
As early as 1637 Rene Descartes
provided and explanation of the rainbow and used Snell's law to study the reflection and refraction of light.
In 1660, Robert Hooke formulated the stress-strain relation of solids and assumed that 
light propagates at a finite speed as oscillations of the medium. In the nineteenth century, Thomas Young was the
first to consider shear as an elastic strain, and in 1809 Etienne Louis Malus discovered polarization of light by reflection. 
In 1821, Fresnel obtained the wave surface of an optically biaxial crystal, assuming that light waves
are vibrations of the ether in which longitudinal vibrations of P waves do not propagate (Carcione and Helbig, 2008). 
He showed that if
light were a transverse wave, then it would be possible to develop a
theory accommodating the polarization of light. 
George Green made extensive use of the analogy between elastic 
waves and light waves, and later James Clerk Maxwell and Lord Kelvin used physical and
mathematical analogies to study wave phenomena in elastic theory and
electromagnetism, when light was discovered to be oscillations of electric and magnetic fields. 
In fact, the displacement current introduced by Maxwell
into the electromagnetic equations arises from the analogy with elastic
displacements.  Maxwell assumed his equations were valid in an absolute
system regarded as a medium (called the ether) that filled the whole
space. The ether was in a state of stress, and would only transmit
transverse waves, but with the advent of the relativity theory the
concept of the ether was abandoned. However, the fact that electromagnetic
waves are transverse waves is important. Carcione and Cavallini (1995) showed that the 2-D Maxwell equations 
describing propagation of the TM mode in anisotropic media                                             
is mathematically equivalent to the SH wave equation in an                             
anisotropic-viscoelastic solid where attenuation is described with the 
Maxwell model.

Any medium (unless vacuum), natural or man-made, dissipates energy when subject to electromagnetic and mechanical fields. 
Dielectric or magnetic relaxation may occur under the influence of electromagnetic fields.  Electric displacement and electric field are related by a relaxation function and a complex dielectric 
permittivity in the time and frequency domains, respectively (B\"ottcher and Bordewijk, 1996; Orfanidis, 2016). 
In viscoelasticity,  stress and strain are related by a relaxation function or a complex stiffness modulus (Nowick and Berry, 1972; Christensen, 1982). 
According to Carcione and Cavallini (1995), one common analogy is given by the fact that 
for dielectric relaxation, electric field strength is equivalent to elastic stress, electric displacement
to elastic strain, and the dielectric constant to elastic compliance. 
For magnetic relaxation, the corresponding variables are the magnetic field strength, intensity of
magnetization, and magnetic susceptibility, respectively. For instance, Carcione and Schoenberg (2000) used these equivalences to introduce viscoelastic relaxation functions and simulate electromagnetic fields. 
However, the mathematical analogy is not complete, since differences arise from the tensor 
nature of stress and strain as compared with the vector nature of the electromagnetic fields.

Proper models should satisfy the Kramers-Kronig relations, known from the beginning of the 20th century from the works of Kronig (1926) and Kramers (1927) on electromagnetism, showing the interrelation between the
real and imaginary parts of the complex susceptibility. 
Electrical and mechanical representations include the Debye model, used to describe the behaviour of dielectric materials, and the Zener viscoelastic model, respectively, both being mathematically equivalent (Carcione, 2014, p. 439). 
In viscoelasticity, the KKRs connect the real and imaginary parts
of the stiffness modulus. Carcione et al. (2019) provide a complete derivation of the relations using the Sokhotski-Plemelj equation, showing explicitly what are the conditions for the relations to hold. 

There are many forms and ways to obtain the KKrs.  Here, two different demonstrations are given, illustrating the eclectic mathematical apparatus available. Then, we apply the acoustic-electromagnetic analogy to obtain the elastic versions. In particular, it is of interest the effects caused by wave propagation, which induces velocity dispersion and attenuation. In this sense, we derive KKrs that related the wave velocity to the attenuation and quality factors. 
Other applications of the KKrs include the magnetic permeability (Silveirinha, 2011), 
circuits (electrochemical impedance) (Esteban and Orazem, 1991), electric circuits (Guillemin, 1949), 
optical spectroscopy (Lucarini et al., 2005) 
and quantum mechanics (Longhi, 2017).
 
\section{The Kramers-Kronig relations}

\subsection{Dielectric permittivity}

Generally, the electric field ${\bf E}$ is defined as the field in vacuum plus a polarization due to the presence of matter. 
The time-domain permittivity is 
\begin{equation} \label{0}
\epsilon (t) = \epsilon_0 H(t) + \chi (t) , 
\end{equation}
where $\epsilon_0$ is the permittivity of vacuum, $\chi$ is the electric susceptibility and $H$ is the Heaviside function.  
The electric displacement is
\begin{equation} \label{1}
{\bf D} = \epsilon \ast \dot {\bf E} = \dot \epsilon \ast {\bf E} = \epsilon_0 {\bf E} + \dot \chi \ast {\bf E} 
\end{equation}
(e.g., Orfanidis, 2016, Section 1.17), since $\dot H = \delta$ (Dirac's function), where ``$\ast$" denotes time convolution and a dot above a variable time differentiation. 
The second term in (\ref{1}) is the polarization. 

Here, we do not make a distinction between the vacuum term and the polarization and consider the permittivity. 
Let us define 
\begin{equation} \label{2}
\hat \epsilon (t) = \epsilon (t) - \epsilon^\infty \ge 0 ,
\end{equation}
such that 
\begin{equation} \label{3}
\epsilon^\infty = \epsilon (t = 0)  \ \ \ {\rm and} \ \ \ \hat \epsilon (t) \rightarrow 0 
 \ \ {\rm when} \ \ t \rightarrow 0 ,
\end{equation}
where $\epsilon^\infty$ is the optical (high-frequency) permittivity.

From equation (\ref{1}), we may write
\begin{equation} \label{a1}
{\bf D} (t) = \int_{-\infty}^t \epsilon (t - t^\prime) \dot {\bf E} (t^\prime) d t^\prime  = 
 - \int_{0}^\infty \epsilon (\tau) \dot {\bf E} (t-\tau) d \tau , 
\end{equation}
because $\epsilon (t)$ is causal and we defined $\tau = t - t^\prime$ (the time derivatives are calculated with respect to the arguments). 
Evidencing explicitly the instantaneous response, 
\begin{equation} \label{a2}
{\bf D} (t) = \epsilon^\infty {\bf E} (t) - \int_{0}^\infty \hat \epsilon (\tau)  \dot {\bf E} (t-\tau) d \tau , 
\end{equation}
where we used equation (\ref{2}). 

Now, substitute a Fourier component for the electric field,
 ${\bf E} (t) = {\bf E}_0 \exp (\ii \omega t)$, to obtain 
\begin{equation} \label{a3}
{\bf D} (t) = \epsilon^\infty {\bf E} (t) + \ii \omega {\bf E}_0 \int_{0}^\infty \hat \epsilon (\tau)  \exp [\ii \omega (t-\tau)] d \tau ,
\end{equation}
where $\ii = \sqrt{-1}$. 

The Fourier transform of equation (\ref{1}) gives
\begin{equation}\label{A1p}
{\cal F} [ {\bf D} ( t ) ] = \epsilon ( \omega ) {\cal F} [ {\bf E} ( t ) ]  , 
\end{equation}
where 
\begin{equation}\label{CMp}
\epsilon ( \omega ) =  {\cal F} [ {\dot \epsilon} ( t ) ] = \int_{- \infty}^\infty \dot \epsilon (t)
\exp ( - \ii  \omega t ) dt
\end{equation}
is the complex permittivity (for clarity, we use the same symbol in both Fourier domains). 

Since 
\begin{equation} \label{a4}
\epsilon (\omega) = \epsilon_1 + \ii \epsilon_2 = {\bf D}/{\bf E} = \epsilon^\infty  + \ii \omega \int_{0}^\infty \hat \epsilon (\tau)  \exp (-\ii \omega \tau) d \tau ,
\end{equation}
we have
\begin{equation} \label{4}
\epsilon_1 (\omega) = \epsilon^\infty + \omega \int_0^\infty \hat \epsilon (t) \sin (\omega t) dt 
\end{equation}
and 
\begin{equation} \label{5}
\epsilon_2 (\omega) =  \omega \int_0^\infty \hat \epsilon (t) \cos (\omega t) dt , 
\end{equation}

Equation (\ref{5}) is a cosine transform, whose reverse transformation is
\begin{equation} \label{7}
\hat \epsilon (t) = \frac{2}{\pi} \int_0^\infty 
\frac{\epsilon_2 (\omega^\prime)}{\omega^\prime} \cos (\omega^\prime t) d \omega^\prime .
\end{equation}

Substituting (\ref{7}) into (\ref{4}) and re-ordering terms yields 
\begin{equation} \label{8}
\epsilon_1 (\omega) = \epsilon^\infty + \frac{2 \omega}{\pi} \int_0^\infty \frac{\epsilon_2 (\omega^\prime)}{\omega^\prime}  \left[
\int_0^\infty \sin( \omega t) \cos (\omega^\prime t) dt \right] d \omega^\prime . 
\end{equation} 
We have 
\begin{equation} \label{9}
\begin{array}{l} 
{ \int}_0^\infty \sin( \omega t) \cos (\omega^\prime t)  dt = 
\myfrac{1}{2} { \int}_0^\infty \left\{ \sin[( \omega + \omega^\prime ) t] + \sin [(\omega - \omega^\prime) t] \right\} dt \\ \\ = 
- \myfrac{1}{2} \left| \frac{\cos[( \omega + \omega^\prime ) t]}{\omega + \omega^\prime} + 
\myfrac{\cos[( \omega - \omega^\prime ) t]}{\omega - \omega^\prime} \right|_0^\infty \\ \\ = 
\myfrac{1}{2} \left[ \myfrac{1}{\omega + \omega^\prime} + 
\myfrac{1}{\omega - \omega^\prime} \right] = \myfrac{\omega}{\omega^2 - {\omega^\prime}^2} ,  
\end{array}
\end{equation} 
where the integrals have been handled as 
\begin{equation} \label{10}
{\rm lim}_{\beta \rightarrow 0}  \int_0^\infty \exp ( - \beta t ) \sin [( \omega \pm \omega^\prime ) t] dt 
\end{equation} 
in the second equality, to warrant convergence at the upper limit  (a more rigorous proof, based on complex-variable theory, is given in $\S$123 of Landau and Lifschitz (1958)],

Then, substituting (\ref{9}) into (\ref{8}) gives
\begin{equation} \label{11}
\epsilon_1 (\omega) = \epsilon^\infty + \frac{2}{\pi} \int_0^\infty \frac{\omega^2 
\epsilon_2  (\omega^\prime)}{\omega^\prime (\omega^2 - {\omega^\prime}^2)}  d \omega^\prime .  
\end{equation} 
Because of the Hermitian properties of $\epsilon (\omega)$
(the permittivity function is real valued) (Bracewell, 2000, p. 16), we have 
\begin{equation} \label{12}
\epsilon_1 (\omega) = \epsilon_1  (-\omega), \ \ \ 
\epsilon_2 (\omega) = - \epsilon_2  (- \omega) , 
\end{equation} 
indicating that $\epsilon_1$ and $\epsilon_2$ are even and odd functions of $\omega$, 
respectively. 

Then, the integrand in (\ref{11}) is an even function and we have 
\begin{equation} \label{13}
\epsilon_1 (\omega) = \epsilon^\infty + \frac{1}{\pi} \int_{- \infty}^\infty \frac{\omega^2 
\epsilon_2  (\omega^\prime)}{\omega^\prime (\omega^2 - {\omega^\prime}^2)}  d \omega^\prime .  
\end{equation} 
Since 
\begin{equation} \label{14}
\frac{\omega^2}{\omega^\prime (\omega^2 - {\omega^\prime}^2)}  = 
 \frac{\omega}{\omega^\prime (\omega - \omega^\prime)} -  \frac{\omega}{\omega^2 - {\omega^\prime}^2} ,  
\end{equation} 
we have 
\begin{equation} \label{15}
\epsilon_1 (\omega) = \epsilon^\infty + \frac{1}{\pi} \int_{- \infty}^\infty \frac{\omega
\epsilon_2  (\omega^\prime)}{\omega^\prime (\omega - \omega^\prime)}  d \omega^\prime  
-  \frac{1}{\pi} \int_{- \infty}^\infty \frac{\omega 
\epsilon_2  (\omega^\prime)}{\omega^2 - {\omega^\prime}^2}  d \omega^\prime .  
\end{equation}
The second integral is zero, because $\epsilon_2$ is odd. Then
\begin{equation} \label{16}
\epsilon_1 (\omega) = \epsilon^\infty + \frac{1}{\pi} \int_{- \infty}^\infty \frac{\omega
\epsilon_2  (\omega^\prime)}{\omega^\prime (\omega - \omega^\prime)}  d \omega^\prime  . 
\end{equation}

We can further simply equation (\ref{13}). Since 
\begin{equation} \label{17}
\frac{\omega}{\omega^\prime (\omega - \omega^\prime )} = \frac{1}{\omega^\prime} + \frac{1}{\omega-\omega^\prime} , 
\end{equation}
we have 
\begin{equation} \label{18}
\epsilon_1 (\omega) = \epsilon^\infty 
+ \frac{1}{\pi} \int_{- \infty}^\infty \frac{
\epsilon_2  (\omega^\prime)}{\omega^\prime}  d \omega^\prime 
+ \frac{1}{\pi} \int_{- \infty}^\infty \frac{
\epsilon_2  (\omega^\prime)}{\omega - \omega^\prime}  d \omega^\prime  .  
\end{equation}
From equation (\ref{7}), the first integral is $\hat \epsilon (t=0)$ since $\epsilon_2 (\omega^\prime)/\omega^\prime$ is an even function. Using equations (\ref{2}) and (\ref{3}), 
\begin{equation} \label{19}
\hat \epsilon (t= 0) = \epsilon (t= 0) - \epsilon^\infty = 0 .
\end{equation}
Then, 
\begin{equation} \label{20}
\epsilon_1 (\omega) = \epsilon^\infty 
+ \frac{1}{\pi} \int_{- \infty}^\infty \frac{
\epsilon_2  (\omega^\prime)}{\omega - \omega^\prime}  d \omega^\prime   
\end{equation}
is another expression of the KK relation 
(B\"ottcher and Bordewijk, 1996, Eq. 8.160).  

Polarization can be dielectric and orientational. The first occurs when a dipole moment is formed in an insulating material due to an externally applied electric field, while 
orientational polarization arises when there is a permanent dipole moment in the material. In B\"ottcher (1993), $\epsilon^\infty$ is the dielectric constant at a sufficiently high frequency, when the permanent
dipoles (i.e. the orientational polarization) can no longer follow the changes of
the field. As stated by B\"ottcher and Bordewijk (1996), equation (\ref{20}) holds for the case that the induced polarization follows the field without delay, implying that the integration should be cut off at a frequency in
the range between the characteristic frequencies of the orientational and the
induced polarization. If the integrations are performed over the whole frequency
range, one implicitly accounts for the fact that the intramolecular
motions connected with the induced polarization are not infinitely fast, and
one should take $\epsilon_0$ instead of $\epsilon^\infty$ in the equation, yielding:
\begin{equation} \label{20p}
\epsilon_1 (\omega) = \epsilon_0 
+ \frac{1}{\pi} \int_{- \infty}^\infty \frac{
\epsilon_2  (\omega^\prime)}{\omega - \omega^\prime}  d \omega^\prime  .
\end{equation}
This corresponds to a KKr for the susceptibility $\chi (\omega ) = \epsilon (\omega ) - \epsilon_0$, according to equation (\ref{0}) (e.g., Orfanidis, 2016, Section 1.17), 
since $\chi_1 = \epsilon_1 - \epsilon_0$ and $\chi_2 = \epsilon_2$.  

Following the same procedure for the imaginary part of the permittivity, we obtain
\begin{equation} \label{21}
\epsilon_2 (\omega) = - \frac{\omega}{\pi} \int_{- \infty}^\infty \frac{
[ \epsilon_1  (\omega^\prime) - \epsilon^\infty]}{\omega^\prime (\omega - \omega^\prime)}  d \omega^\prime  ,
\end{equation}
and the simplified equation is 
\begin{equation} \label{22}
\epsilon_2 (\omega) = - \frac{1}{\pi} \int_{- \infty}^\infty \frac{
\epsilon_1  (\omega^\prime)}{\omega - \omega^\prime}  d \omega^\prime   ,
\end{equation}
since $\epsilon_2 (\omega = \infty)$ = 0.  

Equations (\ref{16})-(\ref{21}) and (\ref{20})-(\ref{22}) are two equivalent Hilbert transform pairs (Bracewell, 2000, p. 359), defining two different expressions of the KKRs. 
The Cauchy principal value of the improper integrals is intended in these calculations. 

Defining the Hilbert transform of a function $g$ as
\begin{equation} \label{HT}
{\cal H} [g (\omega) ] = \frac{1}{\pi} \int_{- \infty}^\infty \frac{g  (\omega^\prime)}{\omega^\prime - \omega}  d \omega^\prime   ,
\end{equation}
equations (\ref{20}) and (\ref{22}) read
\begin{equation} \label{HT1}
\epsilon_1 - \epsilon^\infty = - {\cal H} [\epsilon_2 (\omega) ] 
\ \ \ {\rm and} \ \ \
\epsilon_2 = {\cal H} [\epsilon_1 (\omega) ] .
\end{equation}
respectively

\subsection{Alternative demonstration} 

Since $\hat \epsilon (t)$ [equation (\ref{3})] is real, $\hat \epsilon( \omega )$ is Hermitian; that is
\begin{equation}\label{hermit1}
\hat \epsilon_1 ( \omega ) =  \hat \epsilon_1 ( - \omega ) , \ \ \ \
\hat \epsilon_2 ( \omega ) =  - \hat \epsilon_2 ( - \omega ) .
\end{equation}
Furthermore, $\hat \epsilon$ can split into even and odd functions of time,
$\hat \epsilon_e$ and $\hat \epsilon_o$, respectively, as
\begin{equation}\label{evenodd}
\hat \epsilon (t) = \frac{1}{2} [ \hat \epsilon(t) + \hat \epsilon (-t) ] +
\frac{1}{2} [ \hat \epsilon(t) - \hat \epsilon (-t) ] \equiv \hat \epsilon_e + \hat \epsilon_o .
\end{equation}
Since $\hat \epsilon$ is causal, $\hat \epsilon_o (t) = {\rm sgn}(t) \hat \epsilon_e (t)$, and
\begin{equation}\label{causal}
\hat \epsilon (t) = [ 1 + {\rm sgn} (t) ] \hat \epsilon_e (t),
\end{equation}
whose Fourier transform is
\begin{equation}\label{causal1}
{\cal F} [ \hat \epsilon (t) ] = \hat \epsilon (\omega) 
= \epsilon_1 (\omega ) - \epsilon^\infty - \left( \frac{\ii}{\pi \omega} \right) \ast [\epsilon_1 (\omega) - \epsilon^\infty],
\end{equation}
because ${\cal F} [ \hat \epsilon_e ] = \epsilon_1 - \epsilon^\infty$ and ${\cal F} [ {\rm sgn}(t) ] = 2 /(\ii \omega)$
(Bracewell, 2000, p. 583). 
Equation (\ref{causal1}) and the fact 
that $(\omega - \omega^\prime)^{-1}$ is $\omega$-symmetric and odd 
imply
\begin{equation}\label{KK1}
\epsilon_2 (\omega) = - \left( \frac{1}{\pi \omega} \right) \ast [ \epsilon_1 (\omega) - \epsilon^\infty ]  
= - \frac{1}{\pi} \int_{-\infty}^\infty \frac{ \epsilon_1 ( \omega^\prime ) d \omega^\prime}
{\omega - \omega^\prime} ,
\end{equation}
i.e., equation (\ref{22}) (alternatively, one could also used the property that the Hilbert transform of a constant is zero). 

Similarly, since $\hat \epsilon_e (t) = {\rm sgn}(t) \hat \epsilon_o (t)$, it is 
$\hat \epsilon (t) = [ {\rm sgn} (t) + 1 ] \hat \epsilon_o (t)$, 
and because ${\cal F} [ \hat \epsilon_o ] = \ii \epsilon_2$, we obtain
\begin{equation}\label{KK2}
\epsilon_1 (\omega) - \epsilon^\infty = \left( \frac{1}{\pi \omega} \right) \ast \epsilon_2 (\omega) 
= \frac{1}{\pi} \int_{-\infty}^\infty \frac{ \epsilon_2 ( \omega^\prime ) d \omega^\prime}
{\omega - \omega^\prime} ,
\end{equation}
i.e., equation (\ref{20}). 

In mathematical terms, $\epsilon_1 - \epsilon^\infty$ and $\epsilon_2$ are 
Hilbert transform pairs.
Causality also implies that $\epsilon$ has 
no poles (or is analytic) in the lower half complex $\omega$-plane
(Golden and Graham, 1988, p. 48). A detailed and complete mathematical treatment of the KKrs is given 
in King (2009, Chapter 19).

Including the ionic conductivity into the permittivity infringes the KKrs. 
Carcione (2014, Eq. 8.253) shows that a generalized conductivity can be represented by 
the Kelvin-Voigt model, which does not satisfy the relations, because it does not meet the stability condition (Carcione et al., 2019).
For example, if the medium is conducting at zero frequency,
$\epsilon - \ii \sigma/ \omega$ is singular at $\omega$ = 0. Thus, the KKrs are retained if we subtract this singularity, where $\sigma > 0$ is the zero frequency conductivity. 
Then, the KK-transforms
between the real and imaginary parts of the effective permittivity
are satisfied only if the imaginary part does not contain
any conductivity. This fact has been exploited to retrieve
the conductivity from that effective complex permittivity (Bakry and Klinkenbusch, 2018).  
Expressions of the KKrs for conductors are given in Eqs. (4.8) and (4.9) of Lucarini et al. (2005).

\section{Acoustic-electromagnetic analogy} 

A brief comment on the analogy regarding the KKrs (without equations) is given in Gross (1975). 
The vector analogy between the acoustic and electromagnetic equations, based on the 
Maxwell mechanical model (viscoelasticity) and Maxwell's equations, has been first 
established by Carcione and Cavallini (1995), by which elastic SH waves are mathematically equivalent                                   
to TM (transverse-magnetic) waves. In this particular case, 
\begin{eqnarray}  
\mbox{magnetic field} & \ \  \Leftrightarrow \ \  & \mbox{particle velocity} \nonumber \\                                  
\mbox{electric displacement} & \ \  \Leftrightarrow \ \  & \mbox{elastic strain} \nonumber \\
\mbox{electric field} & \ \  \Leftrightarrow \ \  & \mbox{elastic stress} \nonumber \\            
\mbox{magnetic permeability} \ (\mu) & \ \  \Leftrightarrow \ \  & \mbox{mass density} \ (\rho) \label{e1} \\
\epsilon^\infty & \ \  \Leftrightarrow \ \  & J_U \nonumber \\
\sigma & \ \  \Leftrightarrow \ \  & \eta^{-1} \nonumber \\
\epsilon (\omega) =  \epsilon^\infty - \frac{\ii}{\omega} \sigma 
& \ \  \Leftrightarrow \ \  & 
J (\omega) = J_U - \frac{\ii}{\omega \eta} \nonumber ,
. \end{eqnarray}              
where $\sigma$ is the electrical conductivity, $J$ is the creep compliance, and $J_U$ and $\eta$ are the unrelaxed compliance and viscosity, respectively. 
The compliance may correspond to bulk or to shear deformations.
The equivalence (\ref{e1}) involves also the ionic-conductivity term of Maxwell's equation, which has been included in the permittivity, i.e., we have considered ${\bf D} + \sigma {\bf E}$ instead of ${\bf D}$. 
Other specific analogies are illustrated in Carcione and Robinson (2002), for the reflection-transmission problem, and in Chapter 8 of Carcione (2014). 

For the purpose of this work, we need the following equivalences:
\begin{eqnarray} \label{e2}                                                                 
\epsilon (\omega) & \ \  \Leftrightarrow \ \ & J (\omega) \\
\epsilon^0 & \ \  \Leftrightarrow \ \  & J_R \\
\epsilon^\infty & \ \  \Leftrightarrow \ \  & J_U
, \end{eqnarray}      
where $J_R$ is the relaxed compliance. 

Exploiting the analogy, we easily obtain KKrs equivalent to (\ref{20}) and (\ref{22}):
\begin{equation} \label{C20}
J_1 (\omega) = J_U 
+ \frac{1}{\pi} \int_{- \infty}^\infty \frac{
J_2  (\omega^\prime)}{\omega - \omega^\prime}  d \omega^\prime   = J_U - {\cal H} [J_2 (\omega)] 
\end{equation}
and
\begin{equation} \label{C22}
J_2 (\omega) = - \frac{1}{\pi} \int_{- \infty}^\infty \frac{
J_1  (\omega^\prime)}{\omega - \omega^\prime}  d \omega^\prime = {\cal H} [J_1 (\omega)]  ,
\end{equation}
respectively. These two equations can also be deduced from 
Eqs. (2.4-3) and (2.4-4) of Nowick and Berry (1972) after some calculations, using the property
$\omega^\prime/ ({\omega^\prime}^2- \omega^2) = 
1/(\omega^\prime - \omega)-\omega/({\omega^\prime}^2- \omega^2)$ and the fact that $J_2 (\omega)$ is an odd function. 

Equivalent KKrs can be obtained for the complex stiffness or modulus, $M = M_1 + \ii M_2 = 1/J$, related to the relaxation function: 
\begin{equation} \label{M1}
M_1 (\omega) = M_U 
+ \frac{1}{\pi} \int_{- \infty}^\infty \frac{
M_2  (\omega^\prime)}{\omega - \omega^\prime}  d \omega^\prime   = M_U - {\cal H} [M_2 (\omega)] 
\end{equation}
and 
\begin{equation} \label{M2}
M_2 (\omega) = - \frac{1}{\pi} \int_{- \infty}^\infty \frac{
M_1  (\omega^\prime)}{\omega - \omega^\prime}  d \omega^\prime  = {\cal H} [M_1 (\omega)] .
\end{equation}

These equations have been  shown to be equivalent to Eqs. 2.4-6 and 2.4-7 of Nowick and Berry  (1972) by Carcione and Gurevich (2021). 

\subsection{Wave velocity and attenuation} 

The KKRs can be 
applied to wave velocity and attenuation, which is useful in seismology. 
Let us define the complex wave velocity as 
\begin{equation}\label{K1p}
v_c (\omega ) = \frac{1}{\sqrt{\mu \epsilon (\omega)}} \ \  \Leftrightarrow \ \ \frac{1}{\sqrt{\rho J (\omega)}} = \sqrt{\frac{M (\omega)}{\rho}} ,
\end{equation}
such that the complex slowness is 
\begin{equation}\label{K1}
\frac{1}{v_c}  = \frac{1}{v_p} - \frac{\ii \alpha}{\omega}  ,
\end{equation}
where $v_p$ is the phase velocity and $\alpha$ is the attenuation factor 
(e.g., Carcione, 2014). 

By virtue of the analogy, we may consider the elastic equations (\ref{M1}) and (\ref{M2}). 
Let us identify $1/v_c$ with $M$, 
$1/v_p - 1/v_\infty$ with $M_1 - M_U$ and $- \alpha/\omega$ with $M_2$, where $v_\infty = v_p (\omega = \infty)$ is the unrelaxed velocity. 
Performing similar mathematical developments to obtain those equations, we get
\begin{equation}\label{K2}
\frac{1}{v_p (\omega)} -  \frac{1}{v_\infty} 
= - \frac{1}{\pi}\int_{-\infty}^\infty \frac{ \alpha ( \omega^\prime ) d \omega^\prime}
{\omega^\prime (\omega - \omega^\prime)} 
\end{equation}
and 
\begin{equation}\label{K3}
\alpha (\omega) =   \frac{\omega}{\pi} \int_{-\infty}^\infty 
\frac{d \omega^\prime}
{v_p (\omega^\prime) (\omega - \omega^\prime)} ,
\end{equation}
which are Eqs. (13) in Box 5.8 (p. 169) in Aki and Richards (2009), where we used that 
the Hilbert transform of a constant is zero. These equations hold in the electromagnetic case if 
$v_\infty = 1/ \sqrt{\mu \epsilon^\infty}$ is the optical velocity. 

Using equation (\ref{17}), we easily get 
\begin{equation}\label{K2p}
\omega \left( \frac{1}{v_p (\omega)} -  \frac{1}{v_\infty} \right) 
=  \frac{1}{\pi}\int_{-\infty}^\infty \frac{ \alpha ( \omega^\prime ) d \omega^\prime}
{(\omega^\prime - \omega)} = {\cal H} [\alpha (\omega)] 
\end{equation}
and 
\begin{equation}\label{K3p}
\alpha (\omega) =  - \omega
{\cal H} \left[ \frac{1}{v_p (\omega^\prime)} \right], 
\end{equation}
The first is Eq. (3) in Box 5.8 (p. 167) in Aki and Richards (2009). 

Similar relations between velocity and quality factor ${\cal Q}$ can easily be obtained by considering that 
\begin{equation}\label{K4}
{\cal Q} = \frac{|\omega|}{2 \alpha v_p} ,
\end{equation}
defined as the total energy divided by the dissipated energy on a cycle of wave oscillation
(Carcione, 2014, Eq. 2.124). Then, the KKrs (\ref{K2}) and (\ref{K3}) become
\begin{equation}\label{K5}
\frac{1}{v_p (\omega)} -  \frac{1}{v_\infty} 
= - \frac{1}{2 \pi} \int_{-\infty}^\infty    \frac{|\omega^\prime| }{\omega^\prime v_p (\omega^\prime) {\cal Q} (\omega^\prime)} \frac{d \omega^\prime}
{\omega - \omega^\prime} 
\end{equation}
and
\begin{equation}\label{K6}
\frac{1}{ {\cal Q} (\omega)} = \frac{2 \omega v_p (\omega)}{\pi |\omega|}  \int_{-\infty}^\infty 
\frac{d \omega^\prime}
{v_p (\omega^\prime) (\omega - \omega^\prime)} .
\end{equation}

Carcione et al. (2018) show that the Maxwell and Zener model satisfy the KKRs but the Kelvin-Voigt model do not. If the medium is dispersive but lossless, $\epsilon_1 (\omega) - \epsilon^0$ or  $J_1 ( \omega ) - J_R$ can depend on $\omega$ through functions of $\omega$ whose Hilbert transform involves delta functions, which do not represent damping due to their zero bandwidth. 
In electromagnetism, an specific Lorenz model (e.g., Carcione et al., 2010) satisfies the relations, having zero dissipation in its resonances, where the real permittivity
takes positive and negative values for certain
frequencies even though the imaginary part is zero (Poon and Francis, 2009; Orfanidis, 2016, problem 1.11e). 
On the other hand, lossless dispersion occurs at frequencies is far away from the resonances, where the energy and group velocities coincide (Carcione et al., 2010; Orfanidis, 2016, Sections 1.16 and 1.18). 

An example of dispersionless lossy medium is given by a complex velocity 
$v_c = \omega c / (\omega - \ii \gamma)$, where $c$ is a constant velocity and $\gamma$ a damping factor (Carcione et al., 2016). It is $v_p = c$ and $\alpha = \gamma/c$. However, it can be seen from equation (\ref{K3}) that this medium does not satisfy the KKRs. 

\subsection{Stability and passivity}

The concept of stability is related to the boundedness of the response. A strong condition is square-integrability of $M (\omega )$ along the real axis of the $\omega$-plane, which implies 
\begin{equation}\label{sqiint}
\int_{-\infty}^\infty |M (\omega)|^2 d \omega < C , 
\end{equation}
where $C$ is a  
constant (e.g., Nussenzveig, 1972, p. 95).
This is equivalent to $M ( \omega ) \rightarrow 0$, for 
$|\omega| \rightarrow \infty$ ($\pi \ge {\rm arg} ( \omega ) \ge 0$). 
Generally, this condition cannot be satisfied, but rather 
the weaker one that $|M ( \omega )|$ is bounded, i.e., 
$|M ( \omega )|^2 < C$ is verified. A lossless medium and the Maxwell 
and Zener models satisfy the condition, 
but the Kelvin-Voigt and constant-$Q$ models do not. 
In fact, in the case of the Zener model, $M$ satisfies the weak condition and $M-M(\infty)$ is 
square integrable. A constant-$Q$ model has $M (\omega) \propto \omega^{2 \gamma}$, where $0 < \gamma < 1/2$ (Carcione, 2014, eq. 2.212) and does not satisfy the conditions (Carcione et al., 2019). 
 
Passivity is another characteristic of a causal system by which it can only absorb and
not generate energy. Nussenzveig (1972, App. B) shows that that any linear passive system is causal, so that passivity is a stronger condition. 
In the context of electromagnetism, the condition is 
\begin{equation} \label{pass}
{\rm Im} [ \epsilon (\omega)] < 0, \ \ \ \omega>0 
\end{equation}
(Glasgow et al., 2001) (we use the opposite sign convention for the Fourier transform). For viscoelasticity, the condition is ${\rm Im} [ J (\omega)] < 0$, according to the acoustic-electromagnetic analogy, and  
${\rm Im} [ M (\omega) = 1/J (\omega)] > 0$ (conditions that can be checked with the Zener model (\ref{1aa}) below). 

\section{Numerical evaluation of the Kramers-Kronig relations} 

It is well known that the imaginary part of a complex trace $z$ is the Hilbert transform of its real part (e.g., Cohen, 1995). i.e., 
\begin{equation} \label{N1}
z = z_1 + \ii z_2 = z_1 + \ii {\cal H} [z_1] . 
\end{equation}
Denoting by ${\cal F}$ and ${\cal F}^{-1}$ the forward and inverse Fourier operators, it is 
\begin{equation} \label{N2}
z  = {\cal F}^{-1} [Z] , \ \ \ \
Z = \left\{
\begin{array}{ll}
Z_1  & \mbox{for 0 argument}, \\
2  Z_1 & \mbox{for positive arguments}, \\
0 & \mbox{for negative arguments}, 
\end{array}
\right.
\end{equation}
where $Z_1 = {\cal F} [z_1]$ (Cohen, 1995, p. 30). Then, 
Im [$z$] is the Hilbert transform of $z_1$. 

Let us consider, as an example, equation (\ref{K2p}) to compute 
the phase velocity from the attenuation factor and set $z_1 = \alpha$  and 
$z_2 = \omega (1/v_p - 1/v_\infty)$.  
The work-flow is:

1. Compute $Z_1 = {\cal F} [\alpha]$; 

2. Compute  $Z$ as in equation (\ref{N2}); 

3. Compute  $z  = {\cal F}^{-1} [Z]$; 

4. Set  $\omega \left(\myfrac{1}{v_p} - \myfrac{1}{v_\infty}\right)$ = $z_2$ = ${\cal H} [z_1]$ = Im [$z$]; 

5. Obtain $v_p = \left( \myfrac{1}{v_\infty} + \myfrac{{\rm Im}[z]}{\omega} \right)^{-1}$. 

Let us assume the Zener model, whose complex modulus is
%------ 
\begin{equation} \label{1aa}
M (\omega) = \rho v_\infty^2 \cdot \frac{Q_{0} \left(\sqrt{Q_{0}^2+1}+1 \right)^{-1}+\ii \omega \tau_0} 
{Q_{0} \left(\sqrt{Q_{0}^2+1}-1 \right)^{-1}+\ii \omega \tau_0}, 
\end{equation}
%------  
where $Q_0$ is the minimum quality factor at $\omega_0 = 2 \pi f_0 = 1 / \tau_0$. The high-frequency limit corresponds
to the elastic case, with $M \rightarrow \rho v_\infty^2$. The complex velocity is given by equation 
(\ref{K1p}) and the phase velocity and attenuation factor can be obtained from equation (\ref{K1}).
We assume that the attenuation is known, i.e., 
\begin{equation} \label{att}
\alpha = - \omega {\rm Im} \left[ \frac{1}{v_c} \right]
\end{equation}
and compute $v_p$ using the preceding algorithm. 

%\newpage

\begin{figure}
\includegraphics[width=11cm]{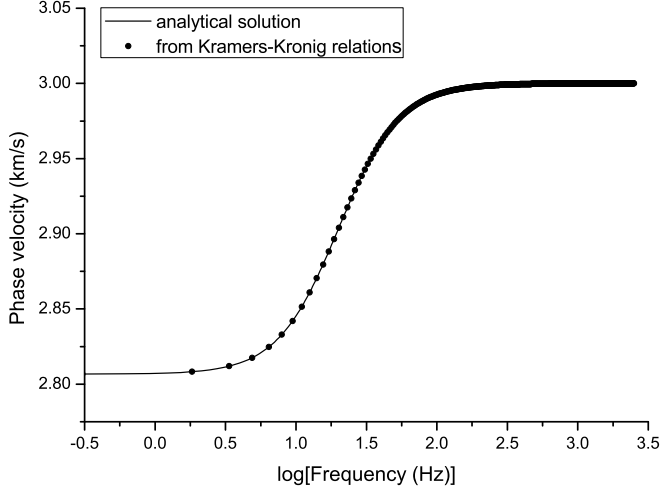}
\caption{Analytical (solid line) and numerical (dots) phase velocities of the Zener (standard-linear solid) model. The latter is computed by numerically solving the 
Kramers-Kronig relations.}
\end{figure}
 
The Fourier transform is computed with the 
fast Fourier transform (FFT) (e.g., Carcione, 2014). The 
arguments (frequencies) are defined as in Eq. 9.70 of Carcione (2014) and the attenuation array is the real part of a complex vector of length equal to a power of two, i.e., $n = 2^{14}$. 
We consider $v_\infty$ = 3 km/s, $Q_0$ = 15 and $f_0$ = 20 Hz. Figure 1 compares the theoretical and numerical phase velocities, where we can observe that the agreement is excellent. 

A generalization of this example to the case of the fractional Zener model can be based on the work of 
Pritz (1999), to analyze the performance of the present algorithm as a function of the order of the fractional derivative.

\end{document}